\documentstyle[psfig]{prhep97}

\makeatletter
\let\chapter\hid@chapter
\makeatother
\begin{document}


\authorrunning{B.\,Klima}
\titlerunning{D\O\ Leptoquark Searches}


\def\talknumber{31304}

\title{Searches for First Generation Leptoquark Pair
Production in $\overline{p}p $\ Collisions at D\O$^1$ }

\author{Boaz\,Klima\
(klima@fnal.gov) (for the D\O\ collaboration)}
\institute{ Fermilab, Batavia, Illinois, U.S.A. }

\maketitle

\footnotetext[1]{Presented at the International Europhysics Conference on High Energy \\
Physics (EPS'97), Jerusalem, Israel, Aug. 19--26, 1997.}


\begin{abstract}
We have searched for the pair production of first generation scalar
leptoquarks using the full data set (123 pb$^{-1}$)
collected with the D\O\ detector at the Fermilab Tevatron during 1992--1996.
We observe no candidates, consistent with the expected  background.
We combine these new results
from the $ee +  {\rm jets}$ and $e\nu + {\rm jets}$
channels with the published $\nu \nu  +{\rm jets}$ result to obtain
a 95\% CL upper limits on the  LQ pair production cross section as a function of
mass and $\beta$, the branching ratio to a charged lepton and a quark. Comparing to the
NLO theory predictions, we set 95\%  CL lower limits on the LQ mass of 225,
204, and 79~GeV/$c^2$ for $\beta=1$, ${1 \over 2}$, and 0, respectively.
The results of this analysis
rule out an interpretation of the excess of high $Q^2$ events at HERA
as leptoquarks with LQ mass below 200 GeV/$c^2$  for  values of $\beta >
0.4$.
\end{abstract}

\section{Introduction}

Leptoquarks (LQ) are hypothesized exotic color-triplet bosons which couple to
both quarks and leptons.  They appear in extended gauge theories and composite
models and  have attributes of both quarks and leptons
such as color,  fractional electric charge, and lepton and baryon quantum
numbers \cite{generic_lq}.
Leptoquarks with universal couplings to all flavors would give rise to
flavor-changing neutral currents and are severely constrained by studies of
low energy phenomena \cite{low_e}.
Therefore, only leptoquarks which couple
within a single generation are considered. The H1 and ZEUS
experiments at
HERA have reported an excess of events at high $Q^2$ in $e^+p$
collisions \cite{hera} \cite{hera_EPS97}.  One possible interpretation
of these events is
production of first generation leptoquarks
at a mass near 200 GeV/$c^2$ \cite{hewett} .

Leptoquarks would  be dominantly  pair-produced via  strong interactions in
$\overline{p}p $\ collisions, independently of  the unknown LQ--$\it l$--$q$
Yukawa coupling.
Each leptoquark would subsequently decay to a lepton and a quark.
For first  generation leptoquarks,  this leads to three possible
final states: $ee+$jets, $e\nu +$jets and $\nu \nu +$jets, with rates
proportional    to   $\beta^2$,     $2\beta(1-\beta)$  and   $(1-\beta)^2$,
respectively, where $\beta$ denotes  the branching fraction of a leptoquark
to an electron and a quark (jet).

Lower limits
on the mass of a first generation leptoquark were published
by the LEP experiments \cite{lep_lq},
by CDF and D\O\ \cite{old_lq}, and by H1 \cite{hera_lq}.

This report describes a search for the pair production of first generation
scalar leptoquarks in the $ee + {\rm jets}$ and $e\nu  +  {\rm   jets}$ final states
using $123 \pm 7$ pb$^{-1}$\ of data collected by D\O\ at the Fermilab Tevatron with
$\sqrt{s} = 1.8$ TeV during 1992--1996. A similar search was conducted by the
CDF collaboration \cite{CDF_EPS97}.
The D\O\ results reported in this article are described in Refs. \cite{d0eeprl}
and \cite{d0enuprl}.
The D\O\ detector and data acquisition system are described in detail in
Ref.~\cite{d0nim}.

\section{The $ee + {\rm jets}$ Channel}

A base data sample of 101 events with two electrons and two or more jets was
selected. The electrons ($E_T^e > 20$~GeV) were required
to be separated from jets ($E_T^j > 15$~GeV).
Events whose
$ee$ invariant mass lies within the
$Z$ boson mass region were rejected.
The efficiency of the trigger used to collect the base
data sample exceeded 99\% for the leptoquark mass range addressed by this
analysis.

Monte Carlo (MC) signal samples were generated for leptoquark masses between
120 and
260 GeV/$c^2$\ using the {\footnotesize{ISAJET}} event generator
and a detector simulation based on the
{\footnotesize{GEANT}} program.
Leptoquark production cross sections were taken from the recently available
next-to-leading order (NLO) calculations~\cite{kraemer}.
The primary backgrounds to the $ee + {\rm jets}$ decay mode are
Drell-Yan+2 jets production (DY),
$t\overline{t}$\ production, and misidentified multijet events.
The {\footnotesize{ISAJET}} DY cross section normalization
was fixed by comparing it with $Z + 2$ jets data.
The D\O\ measured $t\overline{t}$\ production cross section
of $5.5 \pm 1.8$ pb \cite{top_sigma} was used for
top quark MC {\footnotesize{HERWIG}} events.
The multijet background was estimated directly from data \cite{d0eeprl} .

To search for leptoquarks, a random grid search method
was used to optimize cuts on the data and MC samples.
Consistent results were obtained using a neural network \cite{nn}.
The limit setting
criterion of a maximum number of signal events
for a fixed number of background events was adopted. The background
level chosen was 0.4 events, corresponding to a 67\% probability that
no such events would be observed.

The set of cuts which optimally separates signal from background
was determined by a systematic search using distributions of signal
MC events.  Many sets of selection criteria were explored.
A cut on a single, relatively simple variable,
$S_T \equiv H_T^e + H_T^j$, where
$H_T^e \equiv E_T^{e1} + E_T^{e2}$ and
$H_T^j \equiv \sum_{\rm jets} E_T^j$, satisified the limit setting
criterion.
Approximately 0.4 background events are expected for $S_T > 350$ GeV.
No events remain in the base data sample after this $S_T$
cut is applied.
The highest value of $S_T$\ seen in the data is 312 GeV.
The background which is roughly equally
distributed among the three main sources is estimated to be
$0.44 \pm 0.06$ events.

\begin{figure}\vbox{
\centerline{
\psfig{figure=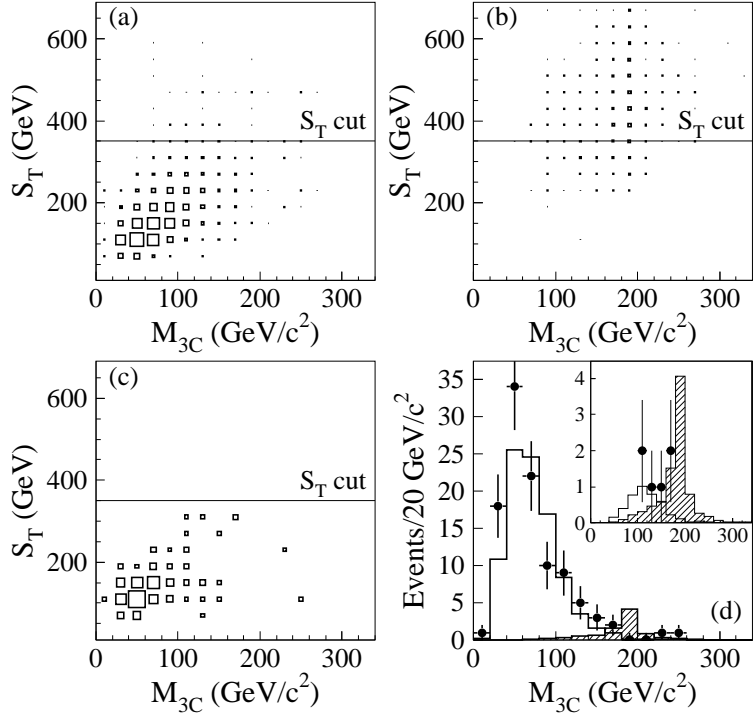,width=4.0in}}
\caption{$S_T$ {\sl vs.}~3C fit mass distributions for (a) background,
(b) 200 GeV/$c^2$\ leptoquarks, and (c) the base data sample.
(d) Mass distribution of the data events
(solid circles), expected background (solid line histogram), and
200 GeV/$c^2$\ leptoquarks (hatched histogram).
The inset plot shows these distributions for $S_T > 250$ GeV.
}
\label{fig:mass_dist}
}
\end{figure}

To investigate the background further, constrained mass fits were
performed on the events in the base data sample, on background samples,
and on the 200 GeV/$c^2$\ leptoquark signal MC sample.
The 3C mass fit was based on the {\footnotesize{SQUAW}}
kinematic mass fitting program and required the two $ej$ masses
to be identical.
Figures~\ref{fig:mass_dist}(a--c) show $S_T$ as a function of the
fit mass for the estimated background,
200 GeV/$c^2$\ leptoquark events, and the base data sample.
The distribution from the data agrees with that of the
expected background.
Figure~\ref{fig:mass_dist}(d) shows the one dimensional
mass distributions for the same samples.
Inset in Fig.~\ref{fig:mass_dist}(d) are the same
distributions after a cut on $S_T > 250$\ GeV.
As can be seen, the data are consistent with the background prediction.

\begin{figure}\vbox{
\centerline{
\psfig{figure=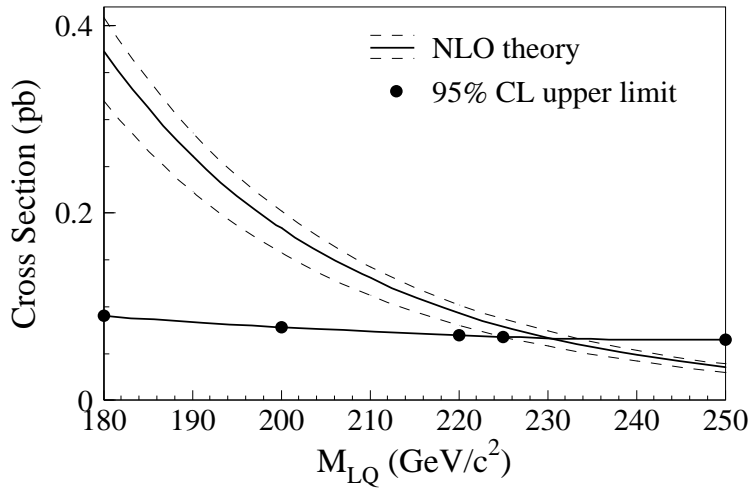,width=4.0in}}
\caption{ Upper limit on the leptoquark pair production cross section
for 100\% decay to $eq$. Also shown is the NLO calculation~\protect\cite{kraemer}
where the central solid line corresponds to
$\mu = M_{LQ}$, and the lower and upper dashed lines to
$\mu = 2M_{LQ}$ and $\mu = M_{LQ}/2$, respectively. }
\label{fig:eelimit}
}
\end{figure}

The overall signal detection efficiency is 9--37\%
for leptoquark masses of 120--250 GeV/$c^2$.
A 95\% confidence level (CL) upper limit on the cross section
$\sigma$ was set using a Bayesian approach
with a flat prior distribution for the signal cross section.
The statistical and systematic uncertainties on the efficiency, the
integrated luminosity, and the
background estimation were included in the limit calculation with
Gaussian prior distributions.  The resulting upper limit on the cross section
is shown in Fig.~\ref{fig:eelimit} together with the NLO calculation of
Ref.~\cite{kraemer}.
The intersection of our limit curve with the lower edge
of the theory band ($\mu = 2M_{LQ}$) is at $\sigma = 0.068$ pb,
leading to a lower limit on the mass
of a first generation scalar leptoquark of 225 GeV/$c^2$.

\section{The $e\nu + {\rm jets}$ Channel}

A base data sample of 14 events with one electron,
two or more jets,
missing $E_T$ ($\rlap{\kern0.25em/}E_T >30$~GeV),
and $M_T^{e\nu}>110$~GeV/$c^2$ was selected.
The  estimated  background is  $17.8 \pm 2.1$
events, of which $11.7 \pm 1.8$, $4.1 \pm 0.9$, and $2.0 \pm 0.7$ events are
from $W+ {\rm jets}$, QCD multijets,  and $\overline{t}t $\ production, respectively.

Monte Carlo samples similar to those in the $ee + {\rm jets}$ were
used in this analysis.
The $W$~+~jets background which is dominant prior to requiring
$M_T^{e\nu}>110$~GeV/$c^2$
was simulated  using the   {\footnotesize{VECBOS}} program.

Two variables  that provide  significant discrimination
between  signal and  remaining  background were identified. They  are the
energy variable $S_T  \equiv E_T^e + E_T^{j1}  + E_T^{j2}
+  \rlap{\kern0.25em/}E_T$, where $E_T^{j1}$
and $E_T^{j2}$ are the  transverse energies of  the two jets, and a
mass variable $\frac{dM}{M}(M_{\rm  LQ})   \equiv {\rm
min}(\frac{|M_{ej}^{(i)}-M_{\rm  LQ}|}  {M_{\rm LQ}};i=1,2)$,
where $M_{LQ}$ is an  assumed leptoquark mass and
$M_{ej}^{(i)}$ are the invariant masses  of the electron with the
two jets.

To find the optimal  selection cuts, the same
criterion as in the $ee + {\rm jets}$ analysis~\cite{d0eeprl}
of maximizing   the signal  efficiency for a  fixed  background of
$\approx$0.4 events was adopted. In the low  mass range ($M_{LQ}  \le $120 GeV/$c^2$ ),
where
the LQ production rates are high, requiring  $S_T >400$~GeV is sufficient.
For $M_{LQ} > 120$ GeV/$c^2$ , neural  networks (NN) were used since they provide
higher  efficiency  than an  $S_T$ cut  alone. At  each mass  where
MC events were generated, a  three layer  feed-forward   neural network
\cite{nn}  was used with two  inputs ($S_T$  and  $\frac{dM}{M}(M_{\rm  LQ})$), five
hidden nodes, and one output node.  Each NN was trained using  simulated LQ
events as the signal (desired  output ${\cal D}_{NN} = 1$) and
a mixture of ~$t\overline{t}$\ ,  $W$~+~jets, and multijet  events as background (with
desired  output  ${\cal  D}_{NN}$~=~0).
Cuts on ${\cal  D}_{NN}$ that yield background estimates in closest
proximity to the desired background were obtained by varying
${\cal  D}_{NN}$ in steps of 0.05.
The expected background  after the  cut  ranges  between $0.29  \pm 0.25$  and
$0.61 \pm 0.27$.
No data events pass the cuts.

\begin{figure}\vbox{
\vspace{-0.25in}
\centerline{
\psfig{figure=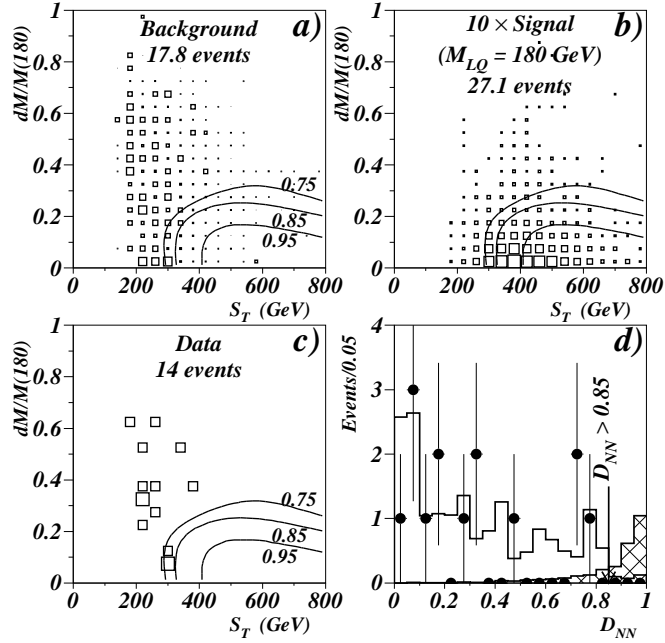,width=4.0in}}
\vspace{-0.15in}
\caption{
$\frac{dM}{M}(180)$ {\sl vs.}   $S_T$    distributions for   (a) predicted
background, (b)  simulated LQ events  ($M_{LQ}$ =  180 GeV/$c^2$), and (c)
data, after  all cuts  except  that on  ${\cal  D}_{NN}(180)$. The contours
correspond to ${\cal  D}_{NN} = 0.75$, 0.85,  and 0.95.  The area of a
box is  proportional to  the number  of events in  the bin,  with the total
number of  events  normalized   to 115  pb$^{-1}$. (d)   Distributions of ${\cal
D}_{NN}$ for  data (solid  circles), background  (solid hist.) and
expected LQ signal for $M_{LQ}$ = 180 GeV/$c^2$ (hatched  hist.).}
\label{fig:nnstdm}}
\end{figure}

Figures~\ref{fig:nnstdm}~(a)-(c)   show the  2-dim. distributions of
$\frac{dM}{M}(180)$ {\sl vs.} $S_T$ for  simulated LQ  signal events with
$M_{LQ}$ =  180 GeV/$c^2$, the  combined   background,  and  data. The
contours  corresponding to constant  values of
${\cal D}_{NN}$
demonstrate  the  separation  achieved between  signal and  background. The
distribution  of ${\cal  D}_{NN}$ for data  is compared  with the predicted
distributions for background and signal in Fig.~\ref{fig:nnstdm}~(d).
The data are described well by background alone. The highest
${\cal  D}_{NN}$ observed in the final data sample is 0.79.

\begin{figure}\vbox{
\vspace{-0.25in}
\centerline{
\psfig{figure=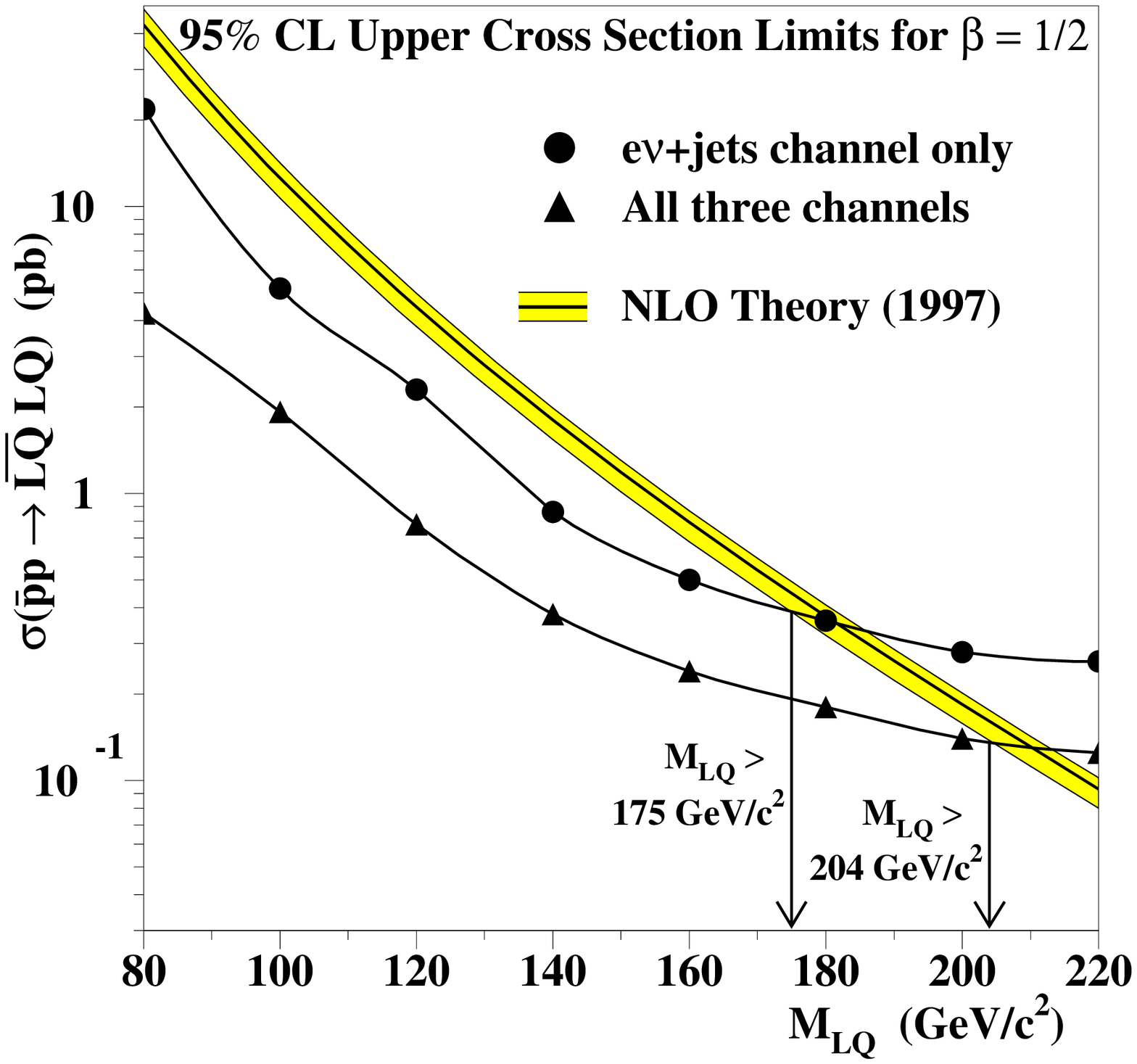,width=4.0in}}
\caption{ Measured 95\%  CL upper limits on the  leptoquark pair production
cross  section in the  $e\nu + {\rm  jets}$  channel (circles) and all
three
channels combined (triangles) for  $\beta= {1 \over 2}$. Also
shown are  the  NLO calculations of   Ref.~\protect\cite{kraemer} where the
central  line corresponds to  $\mu = M_{LQ}$, and the lower
and  upper   lines to   $\mu = 2M_{LQ}$ and  $\mu = M_{LQ}/2$,
respectively. }
\label{fig:enulimit}
}
\end{figure}

Using Bayesian statistics, a  95\% CL upper limit on the leptoquark
pair production cross  section was obtained for $\beta= {1 \over 2}$ as  a function of
leptoquark mass. The statistical and
systematic uncertainties in the  efficiency, the integrated luminosity, and
background  estimation  were  included in  the  limit  calculation with
Gaussian prior probabilities.
  The  measured 95\%  CL cross  section upper limits  for various LQ
masses  are also  plotted in   Fig.~\ref{fig:enulimit}  together  with the NLO
calculations~\cite{kraemer} for  $\beta= {1 \over 2}$. The intersection of the
limit
curve with the lower edge of the  theory band ($\mu = 2M_{LQ}$)
is at $ 0.19$ pb, leading to a 95\% CL lower limit on
the LQ mass of 175 GeV/$c^2$.

\section{Combined Limits}

An     analysis
of  the  $\nu\nu  +  {\rm  jets}$  channel is
accomplished      by   making  use  of   our    published   search
($\int Ldt\approx$7.4  $pb^{-1}$) for the  supersymmetric partner of the top
quark \cite{stop}. Three events survive the selection criteria ($\rlap{\kern0.25em/}E_T >$40
GeV and 2 jets with  $E_T>$30 GeV) consistent  with the  estimated
background of 3.5$\pm$1.2 events. The signal efficiency
for  $M_{LQ}$=  80 GeV/$c^2$ is
calculated   to  be    2.2\%. This
analysis yields the limit $M_{LQ}>$79 GeV/$c^2$ at 95\% CL for $\beta=$0.

\begin{figure}\vbox{
\vspace{-0.2in}
\centerline{
\psfig{figure=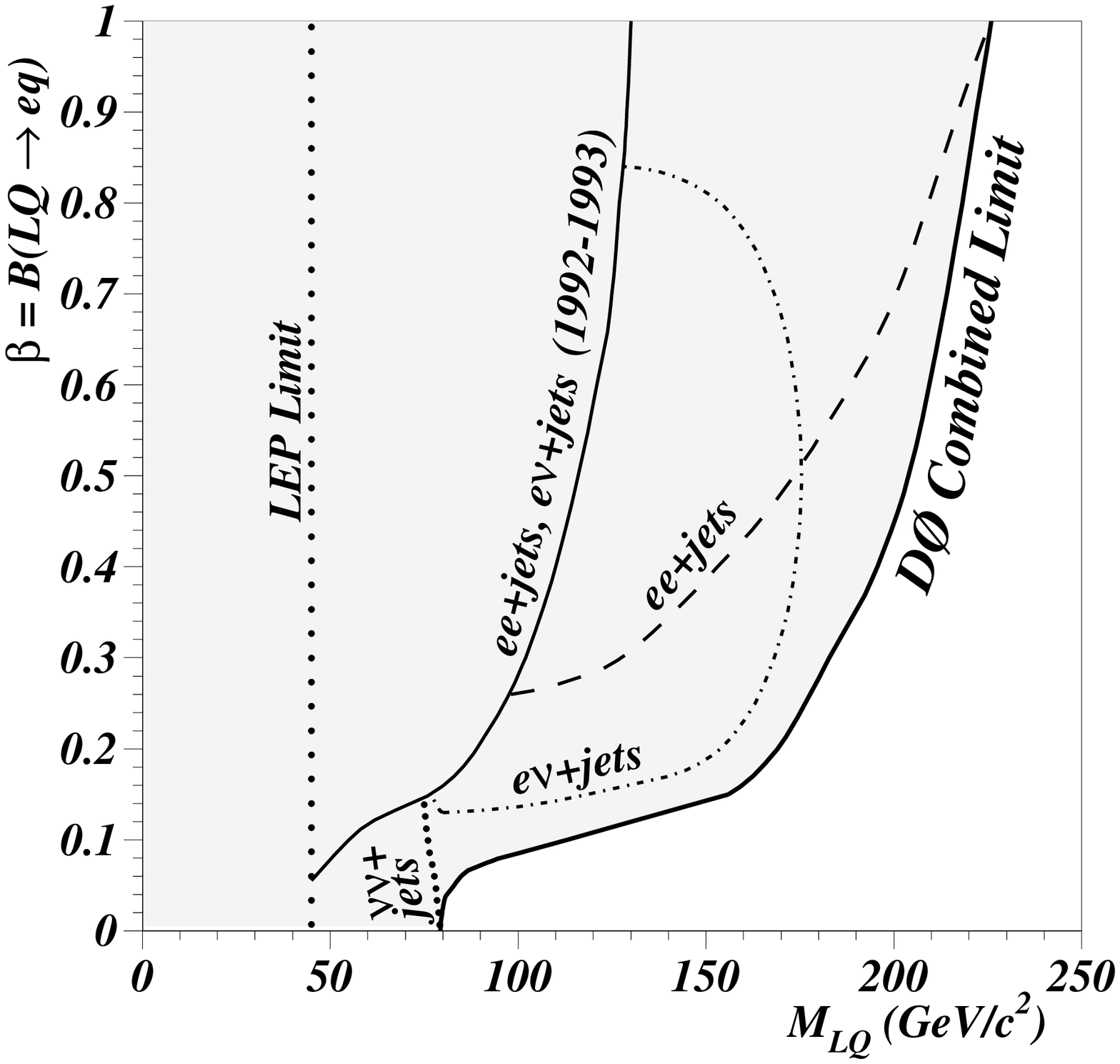,width=4.0in}}
\caption{
Lower limits on the first generation scalar leptoquark mass as a
    function of $\beta$, based on searches in all three possible decay
    channels for leptoquark pairs.  Limits from LEP
    experiments
and from our previous analysis \protect\cite{old_lq} of 1992-93 data are also
shown.
The shaded area is excluded at 95\% CL.
}
\label{fig:exclusion}
}
\end{figure}

Combining the $ee +  {\rm jets}$, $e \nu + {\rm  jets}$, and $\nu\nu + {\rm
jets}$ channels, 95\% CL  upper limits on the LQ pair production cross
section were calculated as a function of LQ mass for various values of
$\beta$.
These cross section limits for  $\beta =  \frac{1}{2}$ (shown in   Fig.
~\ref{fig:enulimit}),
when compared with NLO theory ($\mu=2M_{LQ}$), yield a 95\%
CL lower limit on the LQ mass of 204~GeV/$c^2$.
The lower limits on the LQ mass derived as a function of $\beta$,
from all three channels combined as well as from the individual channels
are shown in   Fig.~\ref{fig:exclusion}. These results
can also  be used to  set  limits on pair   production of any  heavy scalar
particle decaying into a lepton and a quark, in a variety of models.

\section{Conclusion}

We  have  presented a  search for  first  generation scalar
leptoquark pairs  in the $ee + {\rm  jets}$ and $e\nu +  {\rm jets}$
decay  channels. Combining the
results with  those from the  $\nu\nu + {\rm jets}$
channel, we exclude at 95\% CL
 leptoquarks with mass below  225 GeV/$c^2$ for $\beta$ =
1, 204 GeV/$c^2$  for $\beta = {1 \over 2}$, and 79 GeV/$c^2$ for  $\beta = 0$.
Our results  exclude (at 95\%  CL) an interpretation of the HERA high
$Q^2$ event  excess via  $s$-channel LQ  production with LQ mass below 200
GeV/$c^2$ for  values of $\beta > 0.4$.

\section{Acknowledgements}

I would like to express my appreciation to the organizers of this
excellent conference. It is also my pleasure to thank my colleagues
from D\O\ who helped me in preparing this report.


\end{document}